\documentclass[10pt]{iopart}
\usepackage{iopams}         
\usepackage[english]{babel} 
\usepackage{graphicx} 
\usepackage{listings} 
\usepackage{soul}
\usepackage{color}
\usepackage{lipsum}
\usepackage{float}
\usepackage{amsmath}

\begin{document}

\title{Soft colloids for complex interfacial assemblies}

\author{Fabrizio Camerin$^{1,2,3}$  and Emanuela Zaccarelli$^{1,2}$}

\address{(1) CNR-ISC, Sapienza University of Rome, p.le A. Moro 2, 00185 Roma, Italy}
\address{(2) Department of Physics, Sapienza University of Rome, p.le A. Moro 2, 00185 Roma, Italy}
\address{(3) Soft Condensed Matter, Debye Institute for Nanomaterials Science, Utrecht University, Princetonplein 1, 3584 CC Utrecht, The Netherlands}

\ead{fabrizio.camerin@gmail.com}
\ead{emanuela.zaccarelli@cnr.it}
\vspace{10pt}
\begin{indented}
\item[]\today
\end{indented}
%

%
\vspace{2pc}
\noindent{\it Keywords}: colloids $|$ self-assembly $|$ interface

\bigskip

The design of complex materials and the formation of specific patterns often arise from the properties of the individual building blocks. In this respect, colloidal systems offer a unique opportunity because nowadays they can
be synthesized in the laboratory with many different shapes and features. Hence, an appropriate choice of the particle characteristics makes it possible to generate macroscopic structures with desired properties.
The versatility of colloids can also be explored in two dimensions, using liquid-liquid or air-liquid interfaces as privileged substrates
where they can adsorb and self-assemble. Besides being innovative model systems for fundamental studies, the great interest of the scientific community is also technological and applicative, since colloidal-scale surface patterns are very promising for example in photonics or biosensing. In a recent study published in PNAS~\cite{menath2021defined}, Menath and coworkers combine these elements and exploit core-shell colloids, consisting of a silica core and a soft, non-crosslinked polymer shell, to make an important step forward in controlling the assembly of complex structures at an interface.
\\
\\
\indent
One of the main reasons to study colloidal assemblies at interfaces is related to 
the possibility of stabilizing emulsions, in which droplets of a liquid phase are in contact with a second immiscible fluid. In this context, it is worth mentioning Pickering emulsions in which rigid particles coat the interface between the two fluids making the system kinetically stable~\cite{aveyard2003emulsions}.
Soft colloids offer an extra advantage~\cite{style2015adsorption}: being deformable objects, they can spread and flatten at the interface. Their behavior is thus dictated by the combination of the tendency to minimize surface tension, which typically leads to an increase in particle size, and of the elasticity of the object, which counterbalances the previous action. Precisely for these reasons, the conformation that a soft colloid acquires at the interface can be radically different from the one arising in bulk. This has recently started a series of investigations on soft colloids aimed to understand not only the stability of emulsions but also their effective interactions and the structures they form once adsorbed at an interface~\cite{camerin2020microgels,grillo2020self}.

\begin{figure*}[t]
\centering
\includegraphics[width=1\textwidth]{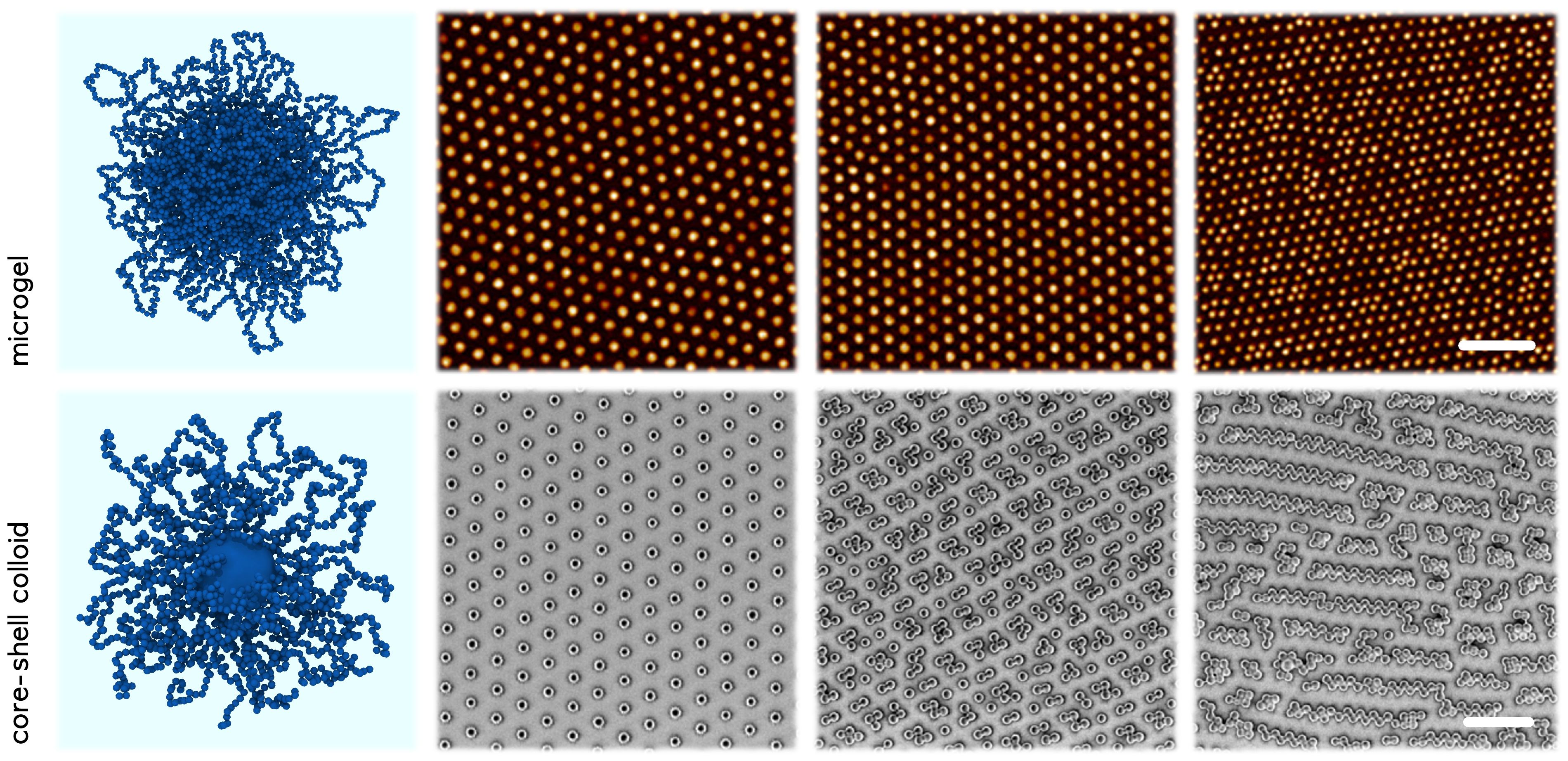}
\caption{\label{fig1}
\textbf{Collective behavior of microgels and core-shell colloids at an interface.} Left: simulation snapshots reporting the top view of a representative microgel (top row) and of a hard core-soft non-crosslinked shell colloid (bottom row) adsorbed at an interface. 		Right: experimental images for microgels (from AFM)~\cite{rey2016fully} and core-shell colloids (from SEM)~\cite{menath2021defined}, respectively. Particles are deposited onto a solid substrate after assembling at an interface for increasing surface pressures (from left to right). While for standard microgels regular hexagonal phases arise, in the case of core-shell colloids unconventional phases with dimers, trimers, tetramers and zig-zag and braided chains are retrieved. Scale bars: 2 $\mu$m for microgels and for core-shell colloids. Adapted with permission from Ref.~\cite{rey2016fully} (Copyright 2016 American Chemical Society) and from Ref.~\cite{menath2021defined} (Copyright 2021 National Academy of Sciences).}
\end{figure*}

In this respect, the case of microgels is emblematic and here serves as a significant case-study, which we will now illustrate for a better understanding of the relevance of the work by Menath \textit{et al.}~\cite{menath2021defined}. Microgels are colloidal-scale polymer networks, that are typically inhomogeneous, being constituted by an inner, more compact core and an outer, fluffier corona. 
One of the first studies about these soft and deformable particles at the interface dates back to about a decade ago~\cite{geisel2012unraveling}. This pioneering work found that the morphology of an individual microgel at a flat water-oil interface resembles that of a ``fried-egg'', as also recently confirmed by computer simulations of a realistic microgel network~\cite{camerin2019microgels}.
In fact the corona, corresponding to the most flexible and less cross-linked portion of the particle, is able to greatly extend at the interface, while the denser core protrudes towards the preferred aqueous phase. This conformation
induces a characteristic collective behavior
which includes, for instance, an isostructural solid-solid transition between two crystalline phases with the same hexagonal symmetry but different lattice constants~\cite{rey2016isostructural,ciarella2021soft}. The intimate link that exists between phase behavior and structure of a single microgel is made particularly evident by this phenomenon. Indeed, following the core-corona structure, upon reaching a certain level of compression, some cross-linked shells may collapse and give rise to a conformation in which core-core contacts are also present. Other studies pointed out the crucial effect of capillary interactions, particularly for large microgels~\cite{scheidegger2017compression} as well as the formation of disordered and glassy states~\cite{camerin2020microgels,huang2017structure}. However, it appears that the typical crystalline arrangement for intermediate compression of microgel particles is limited to the hexagonal one. Typical AFM images illustrating this behavior are shown in the top row of Fig.~\ref{fig1}~\cite{rey2016fully}.  These considerations also apply to similar colloidal particles for which a crosslinked polymer shell surrounds a core of some kind, either soft or hard~\cite{vogel2012ordered,tang2018surface}.

In an effort to achieve structures other than hexagonal, different experimental strategies have been attempted. For instance, the use of a double deposition technique has provided very interesting results and the formation of unconventional structures. In Ref.~\cite{grillo2020self}, Grillo and coworkers described how a first (hexagonal) layer of colloids deposited at the interface acts as a template for a second layer of particles, that are then forced to rearrange forming an overall single layer. In this way, the authors demonstrated that by varying the packing fractions of the two monolayers, it is possible to achieve low-coordinated arrangements as well as structures of higher complexity such as interlocking-S or herringbone superlattices. Similarly, Volk \textit{et al.}~\cite{volk2019moire} showed the formation of Moir\`e and honeycomb lattices from microgels with a rigid core and a soft polymeric shell. 

Altogether this evidence seems to support the idea that starting from building blocks with a substantially spherical shape, it is not 
possible to obtain complex structures from spontaneous assembly. Indeed, the usual approach to direct the assembly towards a certain type of structure is to act on the shape of the colloid so that, by changing interaction potential and packing ability, different structures are generated. It is right in this context that the work of Menath \textit{et al.}~\cite{menath2021defined} is nested. Their idea is rather different: the authors in fact propose to modify the inner structure of the particles rather than its overall shape once adsorbed. The theoretical foundations for this possibility was laid by Jagla~\cite{jagla1998phase} about twenty years ago, who suggested that complex non-hexagonal phases could be achieved for a two-dimensional system with two well-separated length scales. In this way, the outer shells are allowed to shrink, so that other particles can get in core-core contact, once the exerted pressure is sufficiently high. This requirement is actually quite general and does not preclude the formation of conventional hexagonal phases. Indeed, recalling the behavior observed in standard microgels, the presence of two length scales given by the core and the corona leads at most to an isostructural transition between two hexagonal crystalline phases. The authors of the PNAS study therefore point out that more stringent requirements are necessary to advance further, which translates into specific features that the outer shell should retain.

First of all, based on theoretical considerations, the repulsive force exerted by the polymer chains gives rise to a non-convex shape of the colloid-colloid interaction potential. Another important aspect is that each shell should behave independently of the others. The advantage in this case is that the system can minimize its energy by letting compressions to occur in a preferred direction without influencing all the other neighboring particles to behave in the same manner, thus allowing anisotropic structures to form. The major challenge now lies in synthesizing a system that actually meets these conditions. Although the characteristics of the shell may recall systems such as ultra-low-crosslinked~\cite{scotti2019exploring} or hollow microgels~\cite{vialetto2021effect}, these do not represent suitable systems to reproduce this behavior due to the lack of a true internal core. Menath \textit{et al.}~\cite{menath2021defined} propose to use a hard core-soft shell system with clearly separated length-scales, in order to satisfy the first requirement posed by Jagla. In addition, the authors use an advanced radical polymerization scheme to functionalize the core via non-crosslinked polymer chains, being able to obtain shells of highly controlled thickness. In this way, they obtain particles that, once absorbed at the interface, are capable to greatly extend, increasing their size by more than 60\% with respect to their bulk counterpart. As compared to the corona that is typically present in standard microgels, there are some aspects that make this system more eligible to behave as predicted theoretically by Jagla. On one hand, the low-density shell formed by the grafted polymers allows to maintain an almost complete confinement at the interface even at modest compressions, while on the other hand it avoids a marked increase in the overlap of the chains that would lead to a convex shape of the interaction potential. Moreover, the fact that the shell is composed of non-crosslinked polymer chains allows compression of each colloid to be independent of that of its neighbors, thereby satisfying the second aspect that was pointed out in the Jagla conjecture.

By synthesizing the system in this way, it was indeed possible to observe phases other than hexagonal at intermediate compressions of the core-shell colloids at the interface. In fact, a series of cluster phases were observed, with lattices formed first by dimers, then by trimers and finally by tetramers. These in turn give rise to more complex phases such as zig-zag and braided chains, as shown in Fig.~\ref{fig1}. At the maximum possible compression at the interface, larger clusters are formed, in which a great part of the particles are in contact through the cores. The authors then corroborate these observations with Monte Carlo simulations which indicated that a non-convex form of the repulsive interaction potential is needed in order to qualitatively reproduce the phases observed experimentally. 
Hence, we can infer that the peculiarities of the shell are critical for inducing an inversion of the concavity of the effective potential and the related collective behavior for such soft colloids. It would be desirable to find a direct confirmation of this unusual shape of the potential by appropriate monomer-resolved simulations.

The PNAS article by Menath \textit{et al.}~\cite{menath2021defined} therefore provides a fundamental study highlighting the importance of particle softness for the design of new colloidal systems that once confined at an interface are able to assemble in two-dimensions into a variety of different structures. Further strategies will need to be discovered in the near future to go beyond the cluster and zigzag phases observed in this article toward the self-assembly of more complex lattices. To this aim, a combined effort of advanced synthesis, modeling and state-of-the-art simulations and experiments, should be put forward to achieve the desired control of the target structures.

\section*{Acknowledgments}

Our research has been supported by the European Research Council (ERC Consolidator Grant 681597, MIMIC).

\bigskip
\bibliography{biblio}

\providecommand{\newblock}{}
\begin{thebibliography}{10}
\expandafter\ifx\csname url\endcsname\relax
  \def\url#1{{\tt #1}}\fi
\expandafter\ifx\csname urlprefix\endcsname\relax\def\urlprefix{URL }\fi
\providecommand{\eprint}[2][]{\url{#2}}

\bibitem{menath2021defined}
Menath J, Eatson J, Brilmayer R, Andrieu-Brunsen A, Buzza D~M~A and Vogel N
  2021 {\em PNAS\/}

\bibitem{aveyard2003emulsions}
Aveyard R, Binks B~P and Clint J~H 2003 {\em Advances in Colloid and Interface
  Science\/} {\bf 100} 503--546

\bibitem{style2015adsorption}
Style R~W, Isa L and Dufresne E~R 2015 {\em Soft Matter\/} {\bf 11} 7412--7419

\bibitem{camerin2020microgels}
Camerin F, Gnan N, Ruiz-Franco J, Ninarello A, Rovigatti L and Zaccarelli E
  2020 {\em Physical Review X\/} {\bf 10} 031012

\bibitem{grillo2020self}
Grillo F, Fernandez-Rodriguez M~A, Antonopoulou M~N, Gerber D and Isa L 2020
  {\em Nature\/} {\bf 582} 219--224

\bibitem{rey2016fully}
Rey B~M, Elnathan R, Ditcovski R, Geisel K, Zanini M, Fernandez-Rodriguez M~A,
  Naik V~V, Frutiger A, Richtering W, Ellenbogen T {\em et~al.\/} 2016 {\em
  Nano letters\/} {\bf 16} 157--163

\bibitem{geisel2012unraveling}
Geisel K, Isa L and Richtering W 2012 {\em Langmuir\/} {\bf 28} 15770--15776

\bibitem{camerin2019microgels}
Camerin F, Fernandez-Rodr{\'\i}guez M~A, Rovigatti L, Antonopoulou M~N, Gnan N,
  Ninarello A, Isa L and Zaccarelli E 2019 {\em ACS nano\/} {\bf 13} 4548--4559

\bibitem{rey2016isostructural}
Rey M, Fernandez-Rodriguez M~A, Steinacher M, Scheidegger L, Geisel K,
  Richtering W, Squires T~M and Isa L 2016 {\em Soft Matter\/} {\bf 12}
  3545--3557

\bibitem{ciarella2021soft}
Ciarella S, Rey M, Harrer J, Holstein N, Ickler M, Lowen H, Vogel N and Janssen
  L~M 2021 {\em Langmuir\/} {\bf 37} 5364--5375

\bibitem{scheidegger2017compression}
Scheidegger L, Fern{\'a}ndez-Rodr{\'\i}guez M~{\'A}, Geisel K, Zanini M,
  Elnathan R, Richtering W and Isa L 2017 {\em Physical Chemistry Chemical
  Physics\/} {\bf 19} 8671--8680

\bibitem{huang2017structure}
Huang S, Gawlitza K, von Klitzing R, Steffen W and Auernhammer G~K 2017 {\em
  Macromolecules\/} {\bf 50} 3680--3689

\bibitem{vogel2012ordered}
Vogel N, Fernandez-Lopez C, Perez-Juste J, Liz-Marzan L~M, Landfester K and
  Weiss C~K 2012 {\em Langmuir\/} {\bf 28} 8985--8993

\bibitem{tang2018surface}
Tang J~S~J, Bader R~S, Goerlitzer E~S, Wendisch J~F, Bourret G~R, Rey M and
  Vogel N 2018 {\em ACS omega\/} {\bf 3} 12089--12098

\bibitem{volk2019moire}
Volk K, Dei{\ss}enbeck F, Mandal S, L{\"o}wen H and Karg M 2019 {\em Physical
  Chemistry Chemical Physics\/} {\bf 21} 19153--19162

\bibitem{jagla1998phase}
Jagla E 1998 {\em Physical Review E\/} {\bf 58} 1478

\bibitem{scotti2019exploring}
Scotti A, Bochenek S, Brugnoni M, Fernandez-Rodriguez M~A, Schulte M~F, Houston
  J, Gelissen A~P, Potemkin I~I, Isa L and Richtering W 2019 {\em Nature
  communications\/} {\bf 10} 1--8

\bibitem{vialetto2021effect}
Vialetto J, Camerin F, Grillo F, Ramakrishna S~N, Rovigatti L, Zaccarelli E and
  Isa L 2021 {\em ACS nano\/} {\bf 15} 13105--13117

\end{thebibliography}

\end{document}